\documentclass[11pt,twoside]{article}
\usepackage{asp2014}

\aspSuppressVolSlug
\resetcounters

\begin{document}

\title{Stellar population and star formation histories of distant Galactic H II regions NGC 2282 and Sh2-149 complex}
\author{Somnath Dutta$^1$, Soumen Mondal$^1$, Jessy Jose$^2$, R K. Das$^1$
\affil{$^1$S.N. Bose National Centre for Basic Sciences, Kolkata 700106, India; \email{somnath12@boson.bose.res.in}}
\affil{$^2$Kavli Institute for Astronomy and Astrophysics, Peking University,\\
Yi He Yuan Lu 5, Haidian District, Beijing 100871, China;}}

\paperauthor{Somnath Dutta}{somnath12@boson.bose.res.in}{ORCID_Or_Blank}{S.N. Bose National Centre for Basic Sciences}{Department of Astrophysics \& Cosmology}{Kolkata}{West Bengal}{700106}{India}
\paperauthor{Soumen Mondal}{soumen.mondal@bose.res.in}{ORCID_Or_Blank}{S.N. Bose National Centre for Basic Sciences}{Department of Astrophysics \& Cosmology}{Kolkata}{West Bengal}{700106}{India}
\paperauthor{Jessy Jose}{jessyvjose1@gmail.com}{ORCID_Or_Blank}{Kavli Institute for Astronomy and Astrophysics, Peking University}{}{Haidian District}{Beijing}{100871}{China}
\paperauthor{R. K. Das}{ramkrishna.das@bose.res.in}{ORCID_Or_Blank}{S.N. Bose National Centre for Basic Sciences}{Department of Astrophysics \& Cosmology}{Kolkata}{West Bengal}{700106}{India}

\begin{abstract}
We present here the recent results on two distant Galactic H II regions, namely NGC~2282 and Sh2-149, obtained with multiwavelength observations. Our optical spectroscopic analysis of the bright sources have been used to identify the massive members, and to derive the fundamental parameters such as age and distance of these regions. Using IR color-color criteria and H$_\alpha$-emission properties, we have identified and classified the candidate young stellar objects (YSOs) in these regions. The  $^{12}$CO(1-0) continuum maps along with the {\it K}-band extinction maps, and spatial distribution of YSOs are used to investigate the structure and morphology of the molecular cloud associated with these H II regions. Overall analysis of these regions suggests that the star formation occurs at the locations of the denser gas, and  we also find possible evidences of the induced star formation due to the feedback from massive stars to its surrounding molecular medium.
\end{abstract}

\section{Introduction}
Galactic H II regions are the active formation sites of massive O and B type stars.  H II regions are low-density cloud of ionized gas excited by UV radiation from massive stars e.g. O and early B-type \citep{anderson_2014}. H II regions trace active star formations in our galaxy. Galactic H II regions provide us crucial information about the star-formation process, stellar evolution and spiral  arm structures etc. Furthermore, strong stellar winds and UV radiation produced by these massive young stars within an H II region control the structure of their surrounding molecular cloud, and might be possible causes  of triggering new generation of stars or the destruction of the molecular clouds \citep{elemgreen_1977,yorke_1986,koenig_2012}. Thus,  studies of Galactic H II regions containing massive stars provide a platform to understand both high- and low-mass star formations \citep[e.g.][]{jose_2016}.

\begin{figure}
\includegraphics[width=6.0 cm,height=6.0cm]{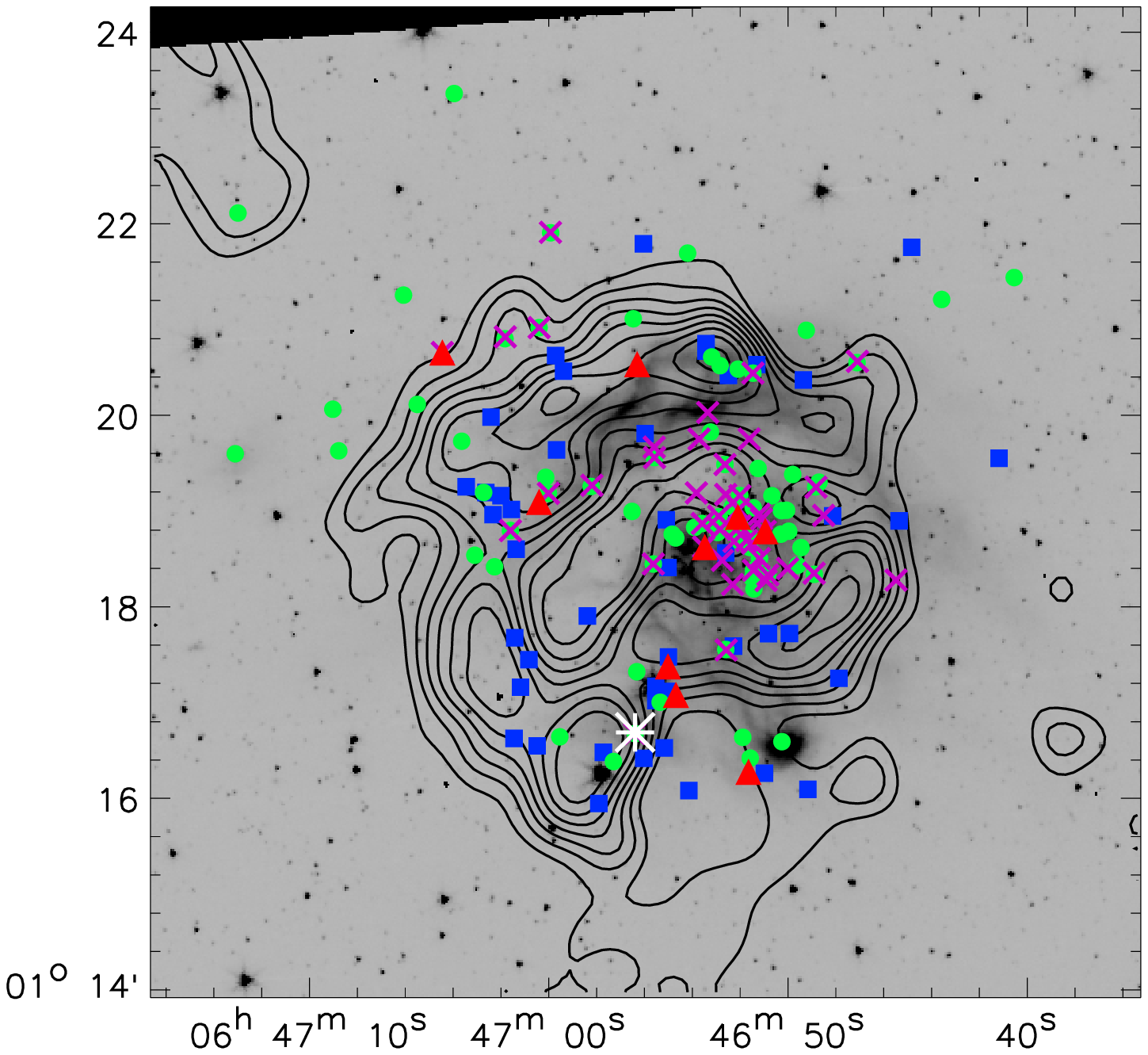}
\includegraphics[width=6.0 cm,height=6.0cm,bb=10 70 620 660]{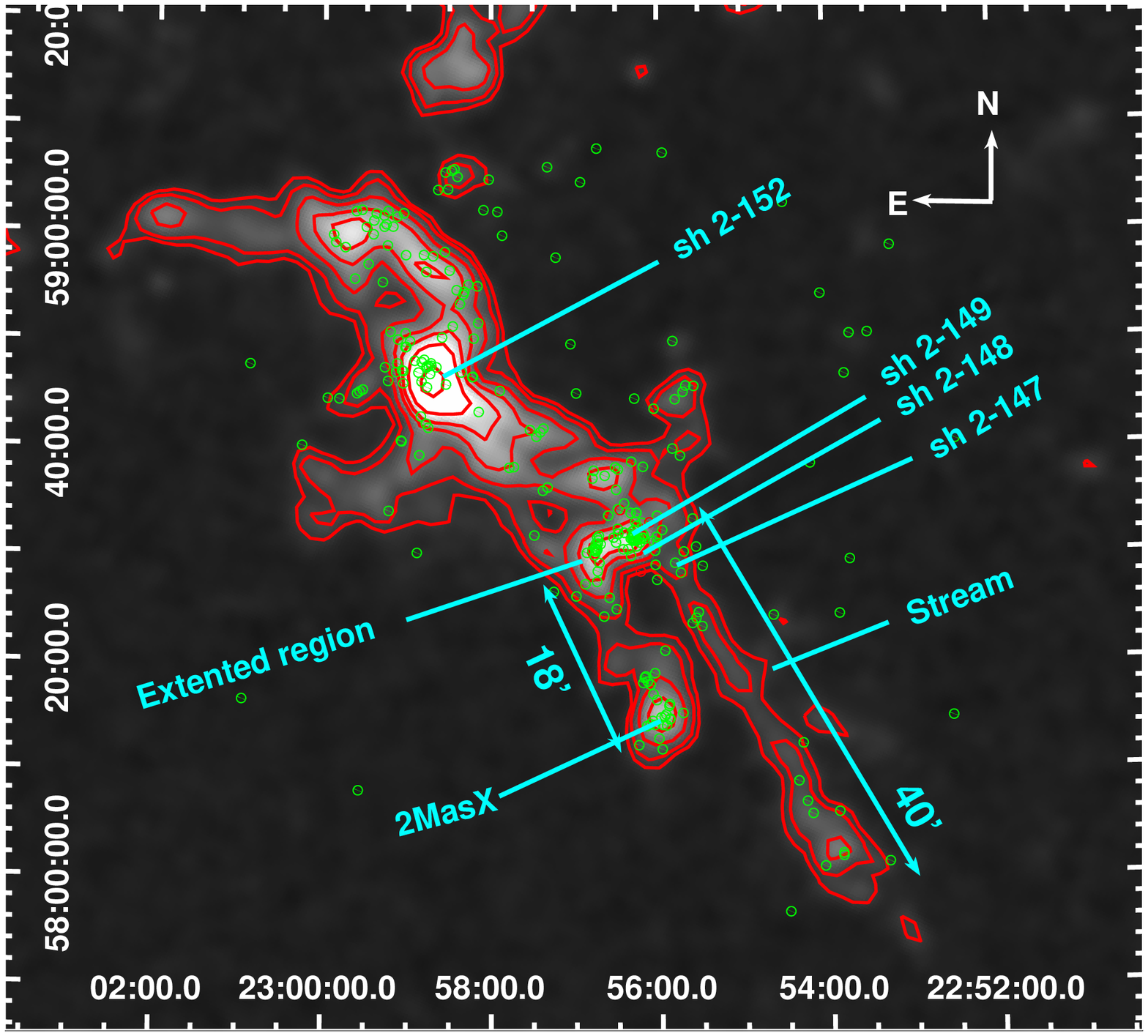}
  \caption{({\it left panel}) Spatial distribution of Class I (red triangles), Class II (green circles), candidate YSOs from $JHK$ colors (blue squares) overlaid on IRAC 3.6 $\mu$m image  of NGC 2282. The magenta crosses indicate the H$_\alpha$ emission line sources. The location of Herbig AeBe star (white asterisk) is also shown. The  contours are plotted for different $A_K$ values from 0.32 to 0.87 mag. ({\it right panel}) $^{12}$CO(1-0) map integrated over -49 km s$^{-1}$ to -54 km s$^{-1}$. Various dense clumps are detected showing evidence of outflow. The contour levels are at 5, 8, 15, 25, 35, 50, 80 K km s$^{-1}$. The locations of Class II (green circles) sources are also marked (see text for details).
} 
  \label{fig:spatial}
\end{figure}

Massive and low-mass young stars in H II regions are embedded in the natal molecular cloud, which are generally invisible at optical wavelengths. However, these regions can be studied in the infrared wavebands. In general, both massive- and low-mass YSOs are formed, and evolved in dense molecular clouds.  Multiwavelength studies of such regions provide census of YSOs, their fundamental parameters e.g. masses, ages, effective temperatures, circumstellar disks around them (if any exists) etc. \citep{kenyon_1995, carpenter_2001, samal_2012, jose_2012}. From such parameter space, broad pictures emerge on the young star-forming regions (SFRs) viz, star-formation history, star-formation efficiency, formation timescales etc. Here, we present the results of the stellar contents of two distant Galactic H II regions e.g. NGC 2282 and Sh2-149 complex using deep optical, near-infrared, mid-infrared data sets and $^{12}$CO(1-0) molecular line observations.

\section{Observations}

For this studies, we have used optical photometry from the 1.04-m Sampurnanand Telescope (ST) of ARIES, Nainital, India and the  2-m Himalayan Chandra Telescope (HCT) of IIA, Hanle, India to estimate ages and masses of YSOs associated with  these regions. Slit spectroscopy of optically bright members and slitless spectroscopy towards the regions were obtained from HFOSC instrument on HCT to estimate the spectral types of the massive stars, distance of the cluster, extinction towards individual star and H$_\alpha$ emission activities of the stars. Issac Newton Telescope (INT) photometric $H_\alpha$ Survey of the Northern Galactic Plane (IPHAS) were conducted using H$_\alpha$, $r$, $i$ filters. We have taken these photometric data from  the archival data release 2 \citep{barentsen_2011}.  We have used  the archive near-infrared J,H,K photometry from  CFHT-WIRCAM \citep{puget_2004},  UKIRT Infrared Deep Sky Survey (UKIDSS, \citep{lawrence_2007}), Two-micron All-Sky Survey (2MASS, \citep{cutri_2003}) point source catalogue, along with mid-infrared photometry from {\it Spitzer}-IRAC to detect the embedded YSOs. \citep[see][for details]{dutta_2015}. Molecular observations in $^{12}$CO(1-0)  from the 14m Five College Radio Astronomy Observatory (FCRAO) Survey of the Outer Galaxy are used to trace  the molecular cloud structure around the H II regions \citep{heyer_1998}.

\section{Results}
\subsection{NGC 2282}

The embedded cluster NGC~2282 ($\alpha_{2000}$ = $06^h46^m50.4^s$ $\delta_{2000}$ = $+01^018^m50^s$ ) is a reflection nebula in the Monoceros constellation located at distance 1.65 kpc \citep[][and the references therein]{dutta_2015}. We have analyzed the stellar surface density distribution of  $K$-band data using nearest neighborhood technique. The radius of the cluster  is  estimated as $\sim$ 3.15$^\prime$ from the semi-major axis of the outer most elliptical contour. From spectrophotometry, we estimated  the spectral types  and membership status of 8 bright sources located  within the cluster area using conspicuous lines and comparison of equivalent widths  with the spectral standards.  Among the bright sources,  three sources are identified as  early B-type members in the cluster, HD 289120, a B2V type star  matches with the previous classification, and two stars (a Herbig Ae/Be star and a B5 V) are newly classified in our study. We estimated the distance to the cluster as $\sim$ 1.65 kpc from spectrophotometric analysis of the massive members. The $K$-band extinction map is estimated from $(H-K)$ color  excess using nearest neighborhood technique, and the mean extinction within the cluster area is found to be $A_V$ $\sim$ 3.9 mag. From slitless spectroscopy, we have identified 16 $H_\alpha$ emission line stars. Another 34 $H_\alpha$ emission line stars are identified from IPHAS  photometry, totaling 50 $H_\alpha$ emission line stars towards the region. Using \citep{gutermuth_2008,gutermuth_2009} scheme, we have classified 9 Class I and 75 Class II objects  using IRAC 3.6 and  4.5 $\mu$m photometry and H and K near-IR photometry. Other candidate YSOs are identified from near-IR $(J-H)/(H-K)$  color-color (CC) diagram.  In total we have identified 152 candidate YSOs using  IR excess and $H_\alpha$ emission properties of the stars towards the region. We characterized these YSOs  using various color-magnitude diagrams (CMD). From $V/(V-I)$ CMD, we have estimated the cluster age which is in the range of $\sim$ 2$-$5 Myr. From mid-IR data, we have estimated the disk fraction of  $\sim$ 58\%, which corresponds to an age of $\sim$ 2-5 Myr. The masses of the candidate YSOs are found to be in the range  $\sim$ 0.1 to 2.0 M$_\odot$ in the  $J/(J-H)$ CMD. The morphology of the region has been studied from spatial distribution of YSOs, stellar density distribution, signature of dust in various optical-infrared images along with the extinction map (Figure~\ref{fig:spatial}. left panel). Our results based on ionizing stars, structure of ionized and molecular gas along with the probable YSOs, though not conclusive, indicate that star-formation activity observed at the border is probably influenced by the expansion of the H II region  to the surrounding medium. More detailed analysis of those results can be obtained from \citep{dutta_2015}.

\subsection{Sh 2-149 complex}

The Sh2-149 ($\alpha_{2000}$ = $22^h56^m16.9^s$ $\delta_{2000}$ = $+58^031^m13^s$) complex located at a distance $\sim$ 5.6 kpc in the constellation Cassiopeia, is  an optically visible region associated with Sh 2-148, Sh 2-147, YSO 2MASX J22555978+5814424 \citep[][and the references therein]{azimlu_2011}. This group of H II regions along with Sh2-152 might be  associated to the supernova remnant SNR G109.10-1.0 \citep{tatamatsu_1985}. We have identified few  optically bright ionizing sources towards  the Sh2-149 complex, and their spectral types of massive O7$-$B0 V are estimated based on optical spectroscopy. We detected several infrared excess stars (candidate YSOs) from NIR and IRAC CC diagrams.  These YSOs are  spatially  distributed along the dust ridge as shown in Figure~\ref{fig:spatial} (right panel), which  indicates that the region is an active star formation site. The morphology from $^{12}$CO(1-0) map identifies various clumps including a stream line flow towards South-West. Mid-IR dust structure suggests that millimeter contours encompass two filamentary structures, one towards East and another towards west, which is an ionized boundary layer towards North-East associated with Sh2-152. The core region of Sh2-149 is extended towards South-East, which is associated to  2MASX J22555978+5814424 via an ionized layer.  The morphology and spatial agreement among millimeter-IR observations, distribution of YSOs and ionizing stars  indicate that the star-formation activity observed at filaments is probably a tentative example of triggered star formation. More detailed analysis will be presented elsewhere (Dutta et al. 2017, in preparation).

\section{Conclusions}

We analyzed the star formation scenario within two distant Galactic H II regions, NGC~2282 and Sh 2-149,  using multiwavelength data sets. We studied the  morphological correlation of gas and dust, nature  of  the ionizing sources and  spatial distribution of  young stellar sources and their properties. The  $^{12}$CO(1-0) maps along with the {\it K}-band extinction maps and spatial distribution of YSOs are used to investigate the structure and morphology of the molecular cloud associated with the H II regions. Our  analysis  also suggests that the star formation occurs at  the locations of the denser gas, and  associate massive members particularly in Sh2-149, might be possible causes of the triggered star formation due to the feedback from massive stars to its surrounding molecular  material. 

\acknowledgements The authors are thankful to the HTAC members and staff of HCT, operated by Indian Institute of Astrophysics (Bangalore); JTAC members and staff of 1.04m ST operated by Aryabhatta Research Institute of Observational Sciences ( Nainital). This research work is financially supported by S N Bose National Centre for Basic Sciences under Department of Science and Technology, Govt. of India. 

\bibliographystyle{asp2014}
\nocite{*}
\bibliography{aspauthor}

\end{document}